\documentclass{article}

\usepackage[final, nonatbib]{neurips_2019_ml4ps}

\usepackage[utf8]{inputenc} % allow utf-8 input
\usepackage[T1]{fontenc}    % use 8-bit T1 fonts
\usepackage{hyperref}       % hyperlinks
\usepackage{url}            % simple URL typesetting
\usepackage{booktabs}       % professional-quality tables
\usepackage{amsfonts}       % blackboard math symbols
\usepackage{nicefrac}       % compact symbols for 1/2, etc.
\usepackage{microtype}      % microtypography
\usepackage{graphicx}

\usepackage{biblatex}

\title{Graph Neural Networks for Particle Reconstruction in High Energy Physics detectors}

\author{
  Xiangyang Ju, Steven Farrell, Paolo Calafiura, Daniel Murnane, Prabhat \\
  Lawrence Berkeley National Laboratory \\
  Berkeley, CA \\
  \texttt{xju@lbl.gov} \\
  \And
  Lindsey Gray,  Thomas Klijnsma, Kevin Pedro, Giuseppe Cerati,\\
  \textbf{Jim Kowalkowski, Gabriel Perdue, Panagiotis Spentzouris, Nhan Tran} \\
  Fermi National Accelerator Laboratory \\
  Batavia, IL \\
  \And
  Jean-Roch Vlimant, Alexander Zlokapa, Joosep Pata, Maria Spiropulu \\
  California Institute of Technology \\
  Pasadena, CA \\
   \And
  Sitong An \\
  CERN, Geneva, Switzerland \& \\
  Carnegie Mellon University, Pittsburgh, PA \\
  \And
  Adam Aurisano, Jeremy Hewes \\
  University of Cincinnati \\
  Cincinnati, OH \\
  \And
  Aristeidis Tsaris \\
  Oak Ridge National Laboratory \\
  Oak Ridge, TN \\
  \And
  Kazuhiro Terao, Tracy Usher \\
  SLAC National Accelerator Laboratory \\
  Menlo Park, CA \\
}

\begin{document}

\maketitle

\begin{abstract}

Pattern recognition problems in high energy physics are notably different from traditional machine learning applications in computer vision. Reconstruction algorithms identify and measure the kinematic properties of particles produced in high energy collisions and recorded with complex detector systems. Two critical applications are the reconstruction of charged particle trajectories in tracking detectors and the reconstruction of particle showers in calorimeters. These two problems have unique challenges and characteristics, but both have high dimensionality, high degree of sparsity, and complex geometric layouts. Graph Neural Networks (GNNs) are a relatively new class of deep learning architectures which can deal with such data effectively, allowing scientists to incorporate domain knowledge in a graph structure and learn powerful representations leveraging that structure to identify patterns of interest. In this work we demonstrate the applicability of GNNs to these two diverse particle reconstruction problems.
%and discuss their advantages relative to traditional algorithmic solutions based solely on domain knowledge.

% Track finding and calorimeter clustering are of vital importance for the High Energy Physics detectors, particularly for the general purposed ones as they have both.
% Traditional algorithms are developed with the help of domain knowledge, resulting in high physics performance but poor scalability.
% This letter demonstrates that the same Graph Neural Networks are good for both track finding and calorimeter clustering.

\end{abstract}

\section{Introduction}
% Introduce the HEP reco problem
The reconstruction of particle collision events in high energy physics experiments such as those at the Large Hadron Collider~\cite{lhc} involves challenging pattern recognition tasks. Particle detectors such as ATLAS~\cite{atlas} and CMS~\cite{cms} are 40m long, 25m diameter instruments with complex geometry and sparse high dimensional data. Specialized detector sub-systems and algorithms are used to reconstruct the different types and properties of particles produced in collisions. For example, charged particle trajectories are reconstructed from spacepoint measurements (``hits'') in tracking detectors, and particle showers are reconstructed from clusters in calorimeters.
The upgraded High-Luminosity LHC~\cite{hllhc}, expected to begin operation in 2026, will deliver increased collision data rates and volumes to the experiments, presenting challenges for current reconstruction solutions.

The traditional approach to particle track reconstruction utilizes combinatorial search algorithms guided by a Kalman Filter. These algorithms are highly tuned for physics performance in today's LHC conditions, but are inherently sequential and scale poorly to the expected HL-LHC conditions with $\mathcal{O}(10^4)$ particles and $\mathcal{O}(10^5)$ hits in each event. The expected challenges of deploying the traditional tracking solutions to HL-LHC data motivated the formation of the HEP.TrkX project to investigate potential new solutions with modern deep learning techniques~\cite{heptrkx-ctd2017, heptrkx-ctd2018}.

% The traditional track-finding algorithm utilizes iterative track-finding 
% seeded from combinations of silicon detector measurements, thus could 
% not be scaled by construction. 
% Moreover, its complexity increases quadratically as a function the number of interactions.
% These drawbacks lead to significant challenges to HEP experiments
% in dense environment such as the High-Luminosity LHC~\cite{hllhc}.
% The HEP.TrkX project~\cite{} explores different modern machine learning architectures,
% including Convolutional Neural Network, Recurrent Neural Network and
% Graph Neural Network (GNN). Among all neural network models, the GNN seems most
% promising in handling the tracking problem.

In this paper we present our work to apply Graph Neural Networks (GNNs) to the particle track and shower reconstruction problems. GNNs were first introduced in \cite{gnn} and have been applied to a growing variety of problems including social networks, knowledge graphs, recommender systems, and 3D shape analysis~\cite{gnn-review1, gnn-review2}. They were first studied for particle tracking applications in \cite{heptrkx-ctd2018} and were also studied for the problem of particle and event classification in \cite{gnn-pileup, gnn-jets, particlenet, gnn-icecube}.

\section{Methodology}

For both tracking and calorimeter cluster problems, we define a graph representation of the data using individual detector measurements as nodes and then constructing edges between nodes with heuristics based on domain knowledge. The GNN models used are based on the Interaction Networks architecture~\cite{interaction-networks}. The primary task of the GNN is to associate detector elements together by classifying the edges of the graph.

\subsection{Tracking}
%GNN was first introduced into the tracking formation in Ref.~[coolpaper]. 
For track finding, we consider only tracks and hits in the barrel region of the detector. The graph is constructed so that the nodes are the hits recorded by the detector and the edges are connections of the hits between adjacent detector layers that pass a pre-defined filter that is tuned to be efficient for tracks resulting from  high transverse momentum particles.
In the input graphs, node features are the three cylindrical coordinates ($r, \phi, z$) and edge features are the difference of the coordinates ($\Delta\eta, \Delta\phi$). The edge labels are 1 if two hits come from the same track, and 0 otherwise.

\begin{figure}[htb]
    \centering
    \includegraphics[width=0.9\textwidth]{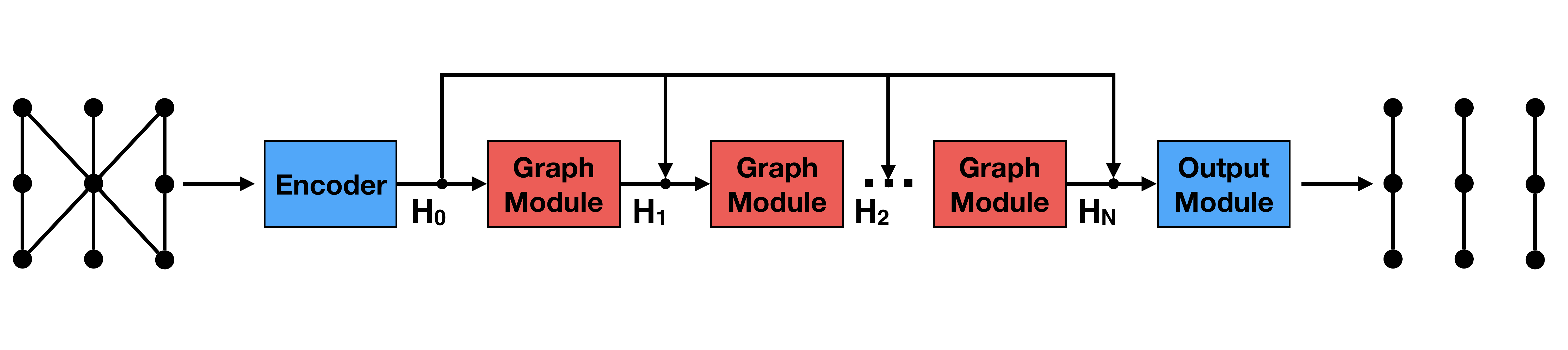}
    \caption{The Graph Neural Network architecture used for tracking.}
    \label{fig:gnn_diagram_tracking}
\end{figure}

% proposed replacement paragraph
The GNN architecture has three components: an encoder which transforms input node and edge features into their latent representations, a graph module which performs message passing to update latent features, and an output module which computes edge classification scores. A diagram of the architecture is shown in figure~\ref{fig:gnn_diagram_tracking}. The encoder uses two fully-connected 2-layer networks for transforming node and edge features, respectively. The initial latent features of the nodes and edges are collectively named $H_0$. The graph module is applied recursively to the latent features. At each iteration $i$ the initial features $H_0$ are concatenated onto the current features $H_i$. This shortcut connection was empirically found to improve model performance. The graph module also uses two fully-connected 2-layer networks, one which computes updated edge features and one which computes updated node features using aggregated incoming edge features. After $N$ iterations of the graph module, the output module takes the last latent features $H_N$ and uses a 2-layer fully-connected network to produce classification scores for every edge. All fully-connected layers use a hidden size of 128 and ReLU activation functions, except the final layer of the output module which uses sigmoid activation. We found that using $N = 8$ graph iterations gave the best model performance.

% The GNN architecture has three components: an input network which transforms input features into latent representations, a graph network which performs message passing to update latent features, and an output network which computes final edge classification scores.
% The input network uses three fully connected layers with hidden and output sizes of 64. The initial latent features of the nodes and edges are collectively named $H_0$.
% The graph network initially takes as inputs the concatenated latent features $[H_i, H_0]$, and then iteratively updates the latent feature $H_i$ via message passing. The last updated $H_i$ is treated as input to the output network that makes the edge classification scores.

\subsection{Calorimeter clustering}

Similar methodologies to those in tracking can be employed to identify energy deposits that should be clustered together to form physically meaningful objects.
In fact, with some minor modifications the same variety of edge classification networks used in the tracking problems described above can be immediately applied to the problem of calorimetry.
If instead of requiring the final output graph to be a collection of tracks, we allow the output graph to be a mesh on a point cloud and label those edges, an energy cluster can be identified.
Moreover, instead of simply being 'true' or 'false' edges the particle type of the edge can also be encoded and inferred. 
This can be achieved with a graph neural network using architectures similar to those demonstrated for tracking as well as networks where the graph is determined dynamically~\cite{calo-gravnet}. Here we will focus on the static graph networks and demonstrate results for future calorimeters in particle physics experiments~\cite{cms-hgcal}.

\begin{figure}[htb]
    \centering
    \includegraphics[width=0.9\textwidth]{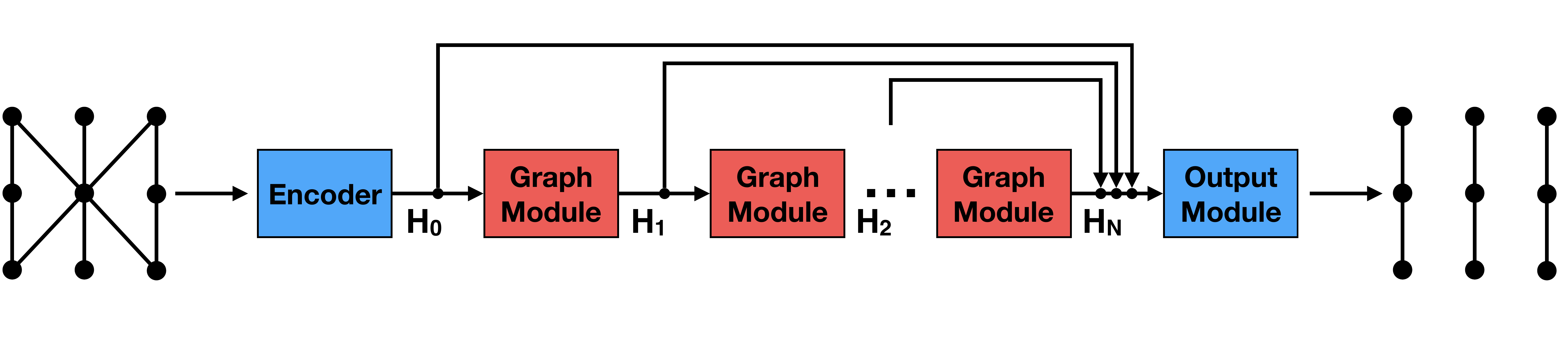}
    \caption{The Graph Neural Network architecture used for calorimeter clustering.}
    \label{fig:gnn_diagram_calorimeter}
\end{figure}

In particular, we have studied the application of message passing networks to the task of calorimeter clustering, yielding initial promising results. The calorimeter clustering problem is very similar to the tracking problem except that there may be more than two true edges connected to an input node. We cast the task of calorimeter clustering as an operation on an initial static graph generated with a simple algorithm like k-Nearest-Neighbours (kNN), passing messages to generate features for classifying those edges as true or false. Here we are using kNN as stand-in for a lightweight reconstruction algorithm as a first pass to generate a graph on the data. The parameter k was chosen such that there was at least one true edge between all hits in the same truth-level cluster after applying the algorithm. Smaller k results in lower clustering efficiency, depending on the use of noise suppression k can be in the range of 8-24.   In particular, these networks use the 'EdgeConv' operator defined in \cite{DGCNN}, and it was found that concatenating the intermediate hidden states in the output stage improved the rate of model convergence by about a factor of two compared to using no such shortcuts.
A diagram of the GNN architecture used for calorimeter clustering is shown in figure~\ref{fig:gnn_diagram_calorimeter}.

\section{Results}

The tracking results are based on the TrackML challenge data~\cite{trkML} generated by the ACTS framework~\cite{acts}. This dataset simulates the very dense environment in the HL-LHC with 200 interactions per bunch crossing on average.

% As a starting point, we focus on the hits recorded in detector Volumes 8, 13 and 17 (i.e. the barrel region) and require there are no missing hits for each track. To form the input graphs, initial edges are the all connections
% between two hits in adjacent layers, then are filtered by tuned selections based on the relative positions of the two hits.
% The efficiency of the edge filter is tuned to be highly efficient for 
% high momentum particles.
% After all these selection, in each event there are about 160000 edges, 
% out of which 92\% are fake.
% From the selected edges direct graphs are constructed as the inputs for GNN. The input node features are the hit cylindrical coordinates, while
% the edge features are the differences of the incoming and outgoing hit coordinates. Due to memory limitations, the input graphs are
% divided into 16 sub-graphs, 8 in $\phi$ bins and 2 in $\eta$ bins,
% which breaks the track that cross multiple graphs.
% Working are going to develop graph algorithms to fully recover
% these discontinued tracks.
% Because the significant imbalance between true edges and 
% fake edges, they are weighted in the loss function, binary entropy.

\begin{figure}[htb]
    \centering
    \includegraphics[width=0.8\textwidth]{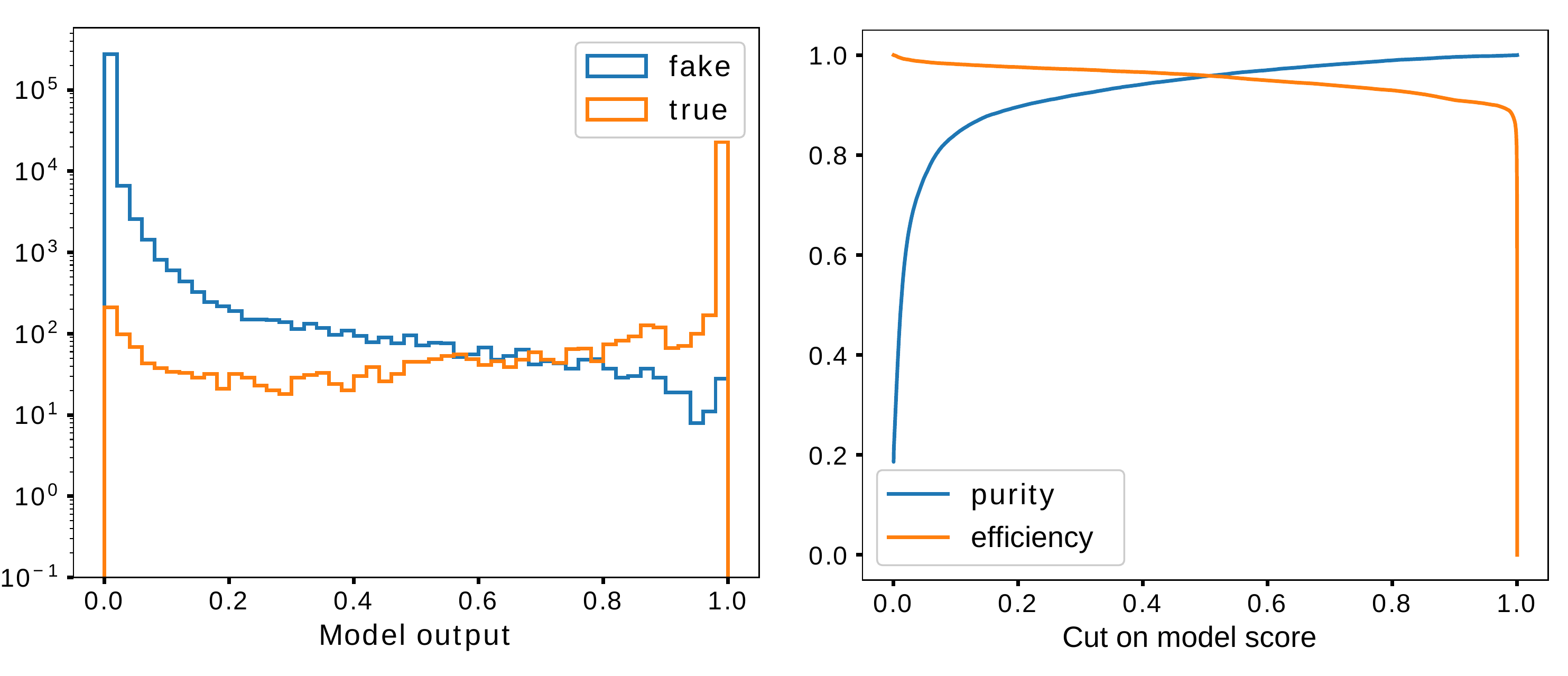}
    \caption{Training results of the Graph Neural Network. Left: The distribution of the edge scores (aka model output) predicted by GNN for true edges that connect the nodes that come from the same track, and for fake edges that do not. Right: The edge purity and efficiency as a function of different cuts on the model score. The definition of efficiency and purity can be found in the text.}
    \label{fig:tracking_roc}
\end{figure}

The GNN is trained on an NVIDIA V100 GPU for about 2 epochs in about two hours, resulting in the performance showed in figure~\ref{fig:tracking_roc}. With a threshold of 0.5 on the GNN output, the edge efficiency, defined as the ratio of the number of 
true edges passing the threshold over the number of total true edges, 
reaches 95.9\%, and the purity, defined as the ratio of the number of
true edges passing the threshold over the number of total edges passing the threshold, is 95.7\%. 
Guided by the GNN outputs, a simple algorithm is used to reconstruct track candidates. The algorithm makes iterative visits to all hits from inside to outside and reconstructs a best track candidate for the hit in question.
Each hit is used only by one track so no ambiguity resolving is needed.
This step is called ``Connecting The Dots'' (CTD).
Using the GNN and CTD together reconstructs about 95\% of true tracks that can be reconstructed in the graph
across the transverse momentum range from 100 MeV to 5 GeV beyond which lacks statistics.
% The Fig.~\ref{fig:tracking_eff} summarizes the fraction of true particles
% after each selection over the total number of tracks in the event for 
% each step.

Ongoing work in reconstructing tracks with GNNs includes extending the method to whole detector data and improving the performance of the CTD post-processing algorithm to recover lost efficiency.

% Our work on track finding shows very promising results.
% Our current focus is to include whole detector data, 
% fully recover the broken tracks. We are also working on
% distributed training of GNN on High Performance Computers, 
% improving single-node performance of the GNN training performance
% and multi-node imbalance.

% \begin{figure}[htb]
%     \centering
%     \includegraphics[width=0.5\textwidth]{selection_efficiency_vs_particle_pt.eps}
%     \caption{The fraction of true particles after each accumulated selection over the total number of tracks in the event as a function of the momentum of the particle ($p_{T}$ GeV).}
%     \label{fig:tracking_eff}
% \end{figure}

\begin{figure}[htb]
    \centering
    \includegraphics[width=0.45\textwidth]{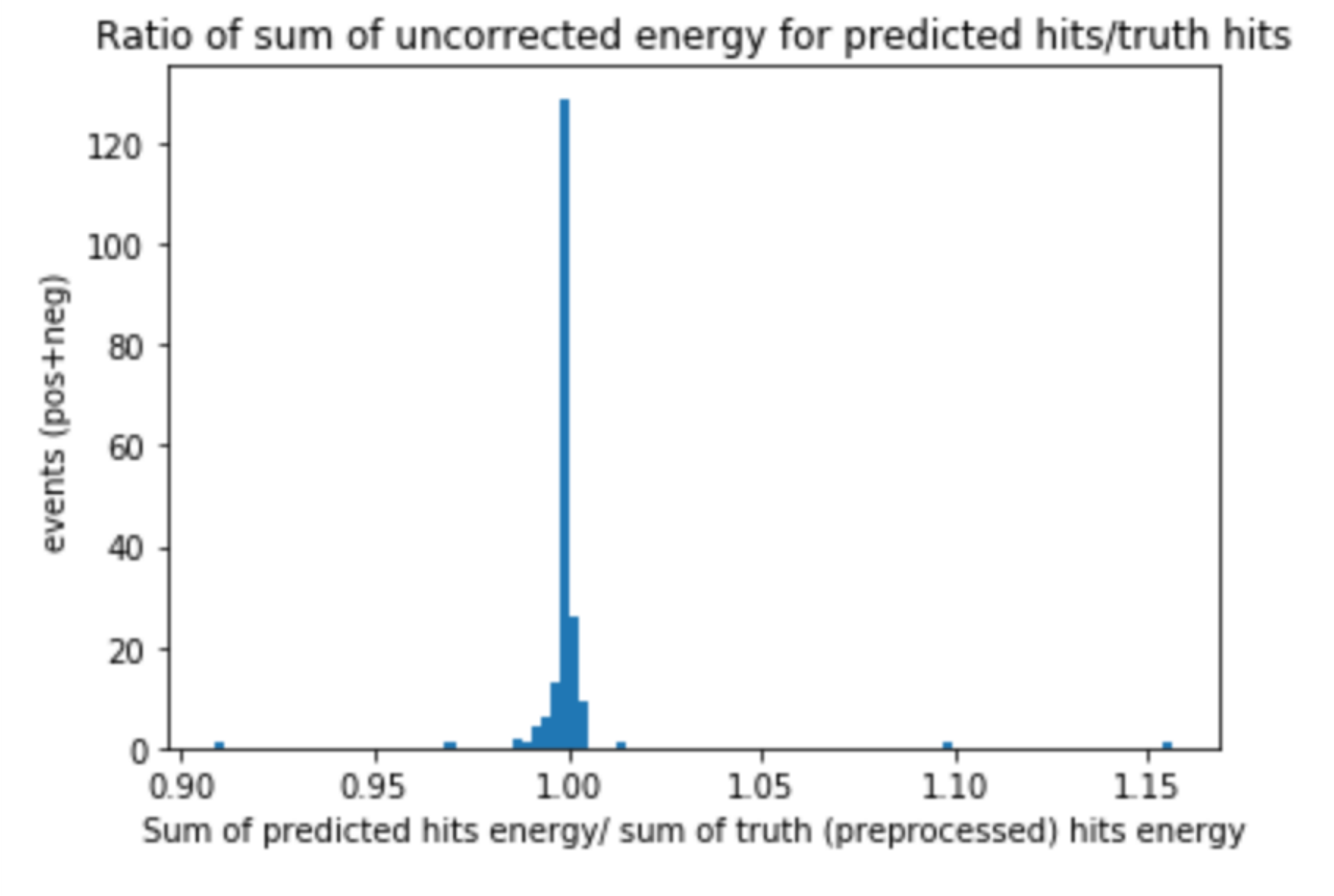}
    \includegraphics[width=0.25\textwidth]{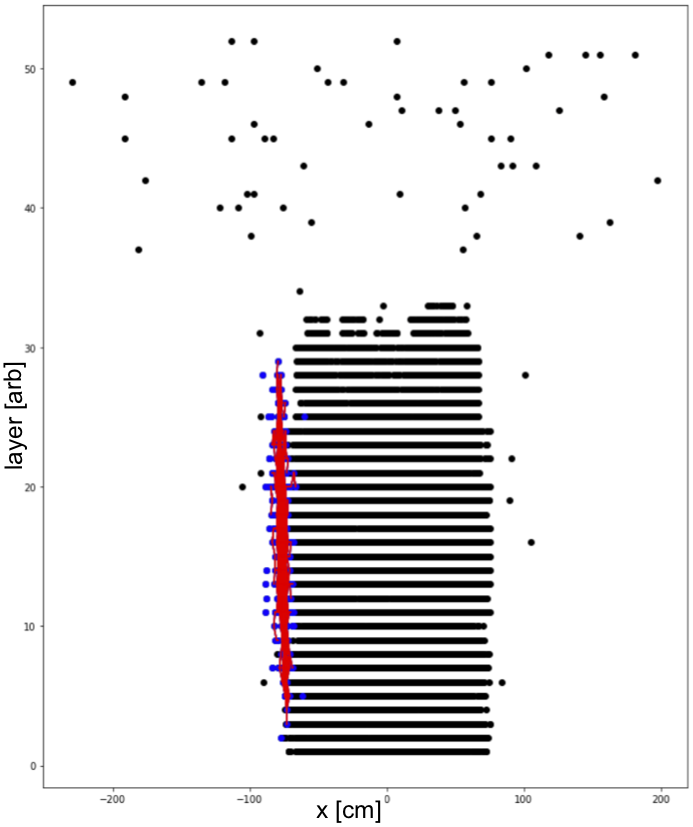}
    \caption{Left: The ratio, per event, for photons of total collected calorimeter energy deposits connected by predicted edges to the energy collected by the associations from ground truth. Right: The event display of a single photon showing the predicted edges (red) and underlying ground-truth nodes (blue) in addition to the energy deposits from noise (black).}
    \label{fig:calo_photon}
\end{figure}

\begin{figure}[htb]
    \centering
    \includegraphics[width=0.8\textwidth]{%
    % graph_c0p98_ievt17.png
    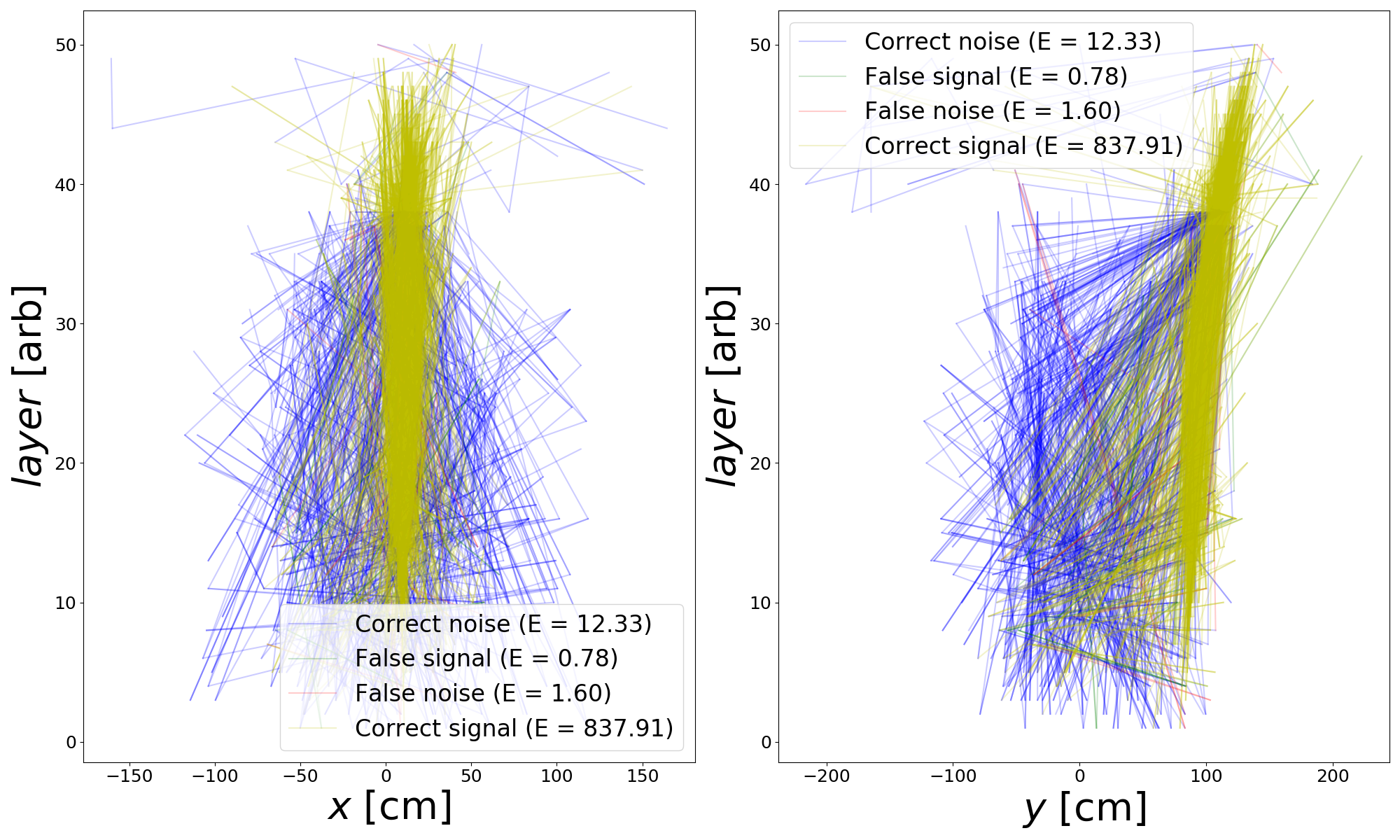%
    }
    \caption{x and y projections of edge classification in a pion shower within the CMS HGCal. The input graph is derived using kNN. The vertical axis in each case is the calorimeter layer number. A score cut of 0.5 is used to identify true edges in this case and edges are labelled according to being true positives (yellow), true negatives (blue), false positives (green), and false negatives (red).}
    \label{fig:calo_pion}
\end{figure}

In the context of calorimetry, we have achieved results separately for muon, photon, and pion energy deposits in the CMS High-Granularity Calorimeter (HGCal). Each variety of particle deposits energy in the calorimeter in a qualitatively different way, with different expected fluctuations in their energy deposition patterns. Pion showers in particular are the most difficult since they exhibit large variability in their shower transverse profile as a function of the shower depth within the calorimeter. In each case, with examples for photon in figure~\ref{fig:calo_photon} and pion in figure~\ref{fig:calo_pion}, we have observed excellent performance for correctly associating energy together using the predictions of these networks. For muons we found 99\% efficiency with 90\% purity, photons we are able to attain 99\% efficiency and purity, and for pions we are able to attain better than 90\% purity and efficiency. The purity of muons is driven by the large amount of noise hits and edges present in the training sample. All of these measurements are made in dedicated single particle samples for each type of particle, the next step of these tests are to move to variable multi-particle final states such as the decays of $\tau$ leptons and then multi-particle jets created in LHC physics events.

This indicates great potential for discovering GNN architectures which can scale to very large number of edges and that can handle multiple high energy particle physics reconstruction tasks. This, in turn, would allow computing centers for high energy physics to focus on certain types of acceleration and better determine where to spend resources and effort in order to become as efficient as possible.

Ongoing work for GNN applications in calorimetry includes studies on how to reconstruct multiple particle types simultaneously using new network architectures which can assign categories to edges. In addition, we are exploring how to better deal with overlapping showers and fractional assignment of calorimeter hit energy into cluster, both of which will be necessary to achieve the best performance for the HGCal. Finally, explorations into deploying these networks for Liquid Argon Time Projection Chambers are in their initial stages.

% \subsection{Ongoing work}

% Our work on track finding shows very promising results.
% Our current focus is to include whole detector data, 
% fully recover the broken tracks. We are also working on
% distributed training of GNN on High Performance Computers, 
% improving single-node performance of the GNN training performance
% and multi-node imbalance.

% For calorimetry, we are studying how to reconstruct multiple particle types simultaneously using new network architectures which can assign categories to edges. In addition, we are exploring how to better deal with overlapping showers and fractional assignment of calorimeter hit energy into cluster, both of which will be necessary to achieve the best performance for the HGCal. Finally, explorations into deploying these networks for Liquid Argon Time Projection Chambers are in their initial stages.

\section{Conclusion}
We have demonstrated that Graph Neural Networks on Point Clouds are suitable for both tracking and calorimetry in high energy physics, having promising physics performance and good scalability.
For the track finding problem, the GNNs combined with a simple connecting-the-dot algorithm results in a relative efficiency of over 95\% for all particles.
Ongoing work is recovering the inefficiency introduced by each selection.
For the calorimeter clustering problem, we have found that very similar graph network architectures yield promising solutions. 
In the individual clustering problems used for testing so far we have found excellent energy collection efficiency, as well as efficiencies and purities better than 90\% even in the most difficult scenarios. 
The next step will be to connect the dots as in the tracking algorithms and derive useful physics quantities from the collections of connected calorimeter energy deposits.

\subsubsection*{Acknowledgments}
We are grateful to Javier Duarte, Phillip Harris, and Jim Hirschauer for the useful discussions.
% Use unnumbered third level headings for the acknowledgments. All acknowledgments
% go at the end of the paper. Do not include acknowledgments in the anonymized
% submission, only in the final paper.
This research was supported in part by the Office of Science, Office of
High Energy Physics, of the US Department of Energy under Contracts No. DE-AC02-05CH11231 and No. DE-AC02-07CH11359, FNAL LDRD 2019.017.
%LG @ PC - DE-AC02-07CH11359 is FNAL for the LDRD

This research used resources of the National Energy Research Scientific Computing Center (NERSC), a U.S. Department of Energy Office of Science User Facility operated under Contract No. DE-AC02-05CH11231.

Part of this work was conducted at "iBanks", the AI GPU cluster at Caltech. We acknowledge NVIDIA, SuperMicro and the Kavli Foundation for their support of "iBanks".

L.G., T.K., K.P., and N.T.~are partially supported by Fermilab LDRD L2019.017: "Graph Neural Networks for Accelerating Calorimetry and Event Reconstruction".

S.A.~is supported by the Marie Sk\l odowska-Curie Innovative Training Network Fellowship of the European Commission’s Horizon 2020 Programme under contract number 765710 INSIGHTS.

\subsubsection*{Software Availability}
The software and the documentation needed to reproduce the results of this article are available at \url{https://github.com/exatrkx/exatrkx-neurips19}

\printbibliography

% \section*{References}

% References follow the acknowledgments. Use unnumbered first-level heading for
% the references. Any choice of citation style is acceptable as long as you are
% consistent. It is permissible to reduce the font size to \verb+small+ (9 point)
% when listing the references. {\bf Remember that you can use more than eight
%   pages as long as the additional pages contain \emph{only} cited references.}

\end{document}